\documentclass[12pt,psf,epsf]{JHEP3}
\usepackage[centertags]{amsmath}
\usepackage{amssymb}
\usepackage{graphicx}
\usepackage{epsfig}
\usepackage{ulem}

\newcommand{\be}{\begin{equation}}
\newcommand{\ee}{\end{equation}}
\newcommand{\bea}{\begin{eqnarray}}
\newcommand{\eea}{\end{eqnarray}}
\newcommand{\N}{{\mathcal N}}
\newcommand{\p}{\partial}

\newcommand{\Y}{\Upsilon}
\newcommand{\BY}{\bar{\Upsilon}}

\newcommand{\bz}{\bar{\zeta}}
\newcommand{\z}{\zeta}
\newcommand{\pp}{\p_{\Upsilon}}
\newcommand{\pb}{\p_{\bar\Upsilon}}
\newcommand{\g}{\mathfrak{g}}
\def\NP{Nucl. Phys.}
\def\PL{Phys. Lett.}
\def\PR{Phys. Rev.}
\def\PRL{Phys. Rev. Lett.}
\def\GRG {Gen. Rel. and Grav.}
\def\CQG {Class. Quant. Grav.}
\newcommand\phm{\phantom{-}}

\title{Critical formation of trapped surfaces in collisions of non-expanding
gravitational shock waves in de Sitter space-time}

\author{I.~Ya.~Aref'eva,$^1$ A.~A.~Bagrov,$^1$ and E.~A.~Guseva$^2$\\
$^1${\it Steklov Mathematical Institute, Gubkin str.~8, 119991, Moscow,
Russia.}\\
$^2${\it Department of Physics, Moscow State University, 119991, Moscow,
Russia.}\\
{ E-mail: arefeva@mi.ras.ru, bagrov.andrey@gmail.com, g14238f@gmail.com}}

\abstract{We study the formation of marginally trapped surfaces
in the head-on collision of two
shock waves in de Sitter space-time as a function of the cosmological constant and the shock wave
energy. We search for a marginally trapped surface on the past light cone of the collision plane. For space-time dimensions $D\ge3$, there exists a critical value of the shock wave energy
above which there is  no trapped surface of this type. For $D>3$, the critical value
of the shock wave energy depends on the de Sitter radius, and there is no this type trapped surface formation
for a large cosmological constant. For $D=3$, the critical value of the shock wave energy is
independent of the cosmological constant. At the critical point, the trapped surface is finite.
Below the critical energy value, the area of the trapped surface depends on the cosmological
constant and the shock wave energy.}


\begin{document}

\newpage

\section{Introduction}

It is a well-established experimental fact that our current universe is
expanding with constant acceleration, which is well described by the
extremely small cosmological constant $\Lambda=10^{-47 }$\,GeV$^4$,
$\Omega_\Lambda=0.726\pm0.015$~\cite{Komatsu:2008hk}. There is a common
opinion that two ultrarelativistic point particles in the Minkowski
space-time can produce black holes~\cite{thooft_s,ACV,AVV}. This
longstanding question has a purely theoretical meaning as well as an
astrophysical one~\cite{Narayan:2005ie}. In the framework of
TeV-gravity~\cite{TeVgravity}, black hole production in collisions of
particles with the center-of-mass energy of a few TeV and their experimental
signatures~\cite{TeVgravity-HEC} became the subject of numerous
investigations~\cite{BF,IA,DL,GT,BH-list1}. We also note a discussion of the
possible production of wormholes and other more exotic objects at the
LHC~\cite{TM,MMT,NS} (see~\cite{INov} for a consideration of wormholes in
astrophysics).

Our main aim in this paper is to determine if the presence of a positive
cosmological constant can influence the processes of black hole formation in
a collision of two ultrarelativistic point particles. Intuitively, it is
almost clear that a small cosmological constant cannot have a detectible
influence. Meanwhile, a positive cosmological constant generates a repulsion
of matter, and a critical value of the cosmological constant $\Lambda$
should exist above which black hole formation is suppressed. Finding this
critical value is worthwhile. We note that under the assumption of an
asymptotically flat space-time, the presence of a trapped surface usually
guarantees the existence of the event
horizon~\cite{Hawking:1969sw,HE,Gibbons:1998zr,Penrose:1999vj}.

We study the formation of marginally trapped surfaces in a head-on collision of two shock waves in
the de Sitter space-time as a function of the cosmological constant and the square of the shock
wave energy. For $D=4$, we show explicitly that there exists a critical value for the ratio of the
shock wave energy $\bar{p}$ and the cosmological radius
\begin{equation}
\label{4cr} \frac{\bar{p}}{a}=\frac{1}{4G}
\end{equation}
above which there is no solution of the trapped surface equation. Here $G$ is the gravitational
constant and the cosmological radius $a$ is related to the cosmological constant by
$a^2=3/\Lambda$. Similar results were found for $D>4$. There is a critical value for the shock wave
energy above which there are no solutions of the trapped surface equation. This value depends on
the cosmological radius. At the critical point, the trapped surface is finite.

This effect is similar to the emergence of critical behavior with respect to
the wave width in the transverse space when a marginally trapped surface is
formed in the head-on collision of two shock waves in the Minkowski
space-time~\cite{AlvarezGaume:2008fx,vw}. This effect depends on the number
of dimensions and is a Choptuik-like critical effect, i.e., it is similar to
the critical Choptuik behavior in gravitational collapse~\cite{choptuikPRL}
(see~\cite{reviews} and the references therein).

The formation of a marginally trapped surface in the collision of
gravitational shock waves in AdS$_D$ was recently studied
in~\cite{gpy,gpy-off-center}. These studies are aimed at better
understanding the entropy production in relativistic heavy ion collisions
due to black hole production in a dual description. Despite the absence of a
holographic dual description of QCD, describing the colliding heavy ions in
terms of colliding gravitational shock waves in the anti-de Sitter
space-time was suggested~\cite{Nastase:2004pc,Nastase:2005rp}. Black hole
formation in collisions of the dual of the nuclei in the bulk is
interpreted as formation of a quark-gluon
plasma~\cite{Shuryak:2005ia,Amsel:2007cw,Grumiller:2008va}. In AdS, a
dimension-dependent critical behavior with respect to the wave width in the
transverse space in the formation of a marginally trapped surface in the
head-on collision of two shock waves was recently
found~\cite{AlvarezGaume:2008fx}. For $D=4$ and $D=5$, there exists a
critical value of this width above which the trapped surface never forms. We
note that in the AdS space-time, the obtained results are qualitatively the
same as those obtained in the flat space-time.

This paper is organized as follows. We start with the setup and recall some
basic facts about the generalization of the Aichelburg--Sexl shock wave
geometry~\cite{aichelburg,thooft,veneziano} to nonexpanding shock waves
propagating in $D$-dimensional space-times with the cosmological
constant~\cite{tanaka,Sfetsos:1994xa,Podolsky:1997ni,Horowitz:1999gf,
Podolsky:2002nn,Podolsky:2001vu,Emparan:2001ce,Kang:2004jd,0606126}. In
Sect.~3, we calculate the critical value of cosmological constant depending
on the shock wave energy below which the trapped surface occurs and give the
area of the trapped surface.

\section{Setup}

Our main aim in this section is to present the setup for studying the
formation of closed trapped surfaces in the head-on collision of two shock
waves in dS space (this may be compared with the setup used to study the
case without a cosmological constant in~\cite{GT} and the case with the
negative cosmological constant
in~\cite{Kang:2004jd,gpy,AlvarezGaume:2008fx}).

We briefly recall the results in~\cite{tanaka,Podolsky:2002nn,0606126} for
the geometry of a shock wave propagating in the $D$-dimensional dS
space-time (see Appendix A). In terms of the dependent plane coordinates,
$(u,v,\vec x)$, $\vec x=(x^2,\dots,x^D)$, satisfying $-2uv+\vec x^2=a^2$
($a$ is the cosmological radius), the line element of the shock wave
space-time is
\be
ds^2=-2du\,dv+d\vec x^2+F(\vec x)\delta(u)du^2.
\ee
The shock wave shape function $F$ is a fundamental solution of the equation
\be
\left(\triangle_{\mathbb{S}^{D-2}}+\frac{D-2}{a^2}\right)F=
-16\sqrt{2}\pi G_D\bar p\delta(\vec n,\vec n_0),
\label{ee1}
\ee
where $\triangle_{\mathbb{S}^{D-2}}$ is the Laplace--Beltrami operator on a
$(D{-}2)$-dimensional sphere $\mathbb{S}^{D-2}$,
$\vec n=\vec x/|\vec x|$, $\vec n_0$ is the location of the particle
on the sphere, $\bar p$ is the energy of the shock wave, and $G_D$ is the
$D$-dimensional gravitational constant. The $\sqrt{2}$ in the right-hand
side results from our choice of the coordinate system (if we impose
$-du\,dv$ instead $-2du\,dv$ in the expression for the linear element, then
it disappears). This metric is a solution of the Einstein equations for an
energy-momentum tensor with the single nonvanishing component
$T_{uu}\sim\bar p\delta(u)$. In the standard parameterization of the
$(D{-}2)$-dimensional sphere by spherical angles
$\vartheta_1,\dots,\vartheta_{D-2}$, the shock wave shape function
corresponding to an ultrarelativistic point particle depends on only one
spherical angle $\vartheta_{D-2}$. The operator
$\triangle_{\mathbb{S}^{D-2}}$ acts on $F=F(\vartheta_{D-2})$ as
\be
\triangle_{\mathbb{S}^{D-2}}F=\frac{1}{a^2}
\sin^{3-D}(\vartheta_{D-2})\left(\frac{d}{d\vartheta_{D-2}}
\sin^{D-3}(\vartheta_{D-2})\frac{d}{d\vartheta_{D-2}}\right)F(\vartheta_{D-2})
\ee
(see~\cite{Podolsky:1997ni} regarding shock waves in dS/AdS with multipole
structures).

For $D=4$, we deal with
\be
\label{eq:F4}
F(\xi)=4\sqrt2p\left(-2+\xi\ln\left(\frac{1+\xi}{1-\xi}\right)\right),
\ee
where $\xi=x^4/a=\cos\vartheta _2$ and $p=\bar p G_4$ is the rescaled
energy.

We now consider a collision of two waves of the type described above. We
suppose that in the region $\{u<0\}\cup\{v<0\}$, i.e., the part of the
space-time before the collision, the metric is given by
\be
ds^2=-2du\,dv+d\vec x^2+F(\xi,\xi_1)\delta(u)du^2+
F(\xi,\xi_2)\delta(v)dv^2,
\label{two-sw}
\ee
where $\xi_1$ and $\xi_2$ are the locations of the two colliding particles
(see~\cite{tanaka} for the explicit formula for $F(\xi,\xi_i)$). In
independent coordinates (see Appendix~C.1), the metric is
\be
ds^2=\frac{-2dw\,d\sigma+2d\z\,d\bz+
2H_1(\z,\bz)\,\delta(w)dw^2+
2H_2(\z,\bz)\,\delta(\sigma)d\sigma^2}
{[1-(w\sigma-\z\bz)/2a^2]^2},
\label{two-sw-ind}
\ee
where $H_i(\z,\bz)=H(\z,\bz,\z_i,\bz_i)$,
$\z_i=\z(\xi_i,\bar\xi_i)$, and
\be
\label{eq:complex-sw-s}
H(\z,\bz,0,0)=H(\z\bz)=\frac12
\left(1+\frac{1}{2a^2}\z\bz\right)
F\left(\frac{1-\z\bz/2a^2}{1+\z\bz/2a^2}\right).
\ee

A rigorous analysis of the formation of black holes in collisions would
require solving the Einstein equations in the interaction region
$\{w>0,\sigma>0\}$ (see, e.g.,~\cite{Gr,AVV} and the references therein).

A sufficient condition for a black hole to form in the asymptotically flat
case is the existence of a marginally closed trapped surface at the
hypersurface $\{w\le0,\sigma=0\}\cup\{w=0,\sigma\le0\}$~\cite{penrose,dp,
GT,Kang:2004jd,gpy}. We note that in non-asymptotically flat cases, there
are no general theorems, but there is a common opinion that the existence of
a marginally trapped surface can be used as an indication of black hole
formation.

In the coordinates used in line element~\eqref{two-sw-ind} null geodesics are discontinuous across
the wave fronts, $w=0$ and $\sigma=0$ (see Appendix C) . So using such coordinates to find this
trapped surface equation is inconvenient and can be avoided by switching to a new coordinate system
$(W,\Sigma,\Y,\BY)$ in which the delta function terms are eliminated in the metric and the
geodesics are continuous. Similar to the D'Eath and Payne coordinates~\cite{dp}, which are closely
related to the explicit form of geodesics in the Minkowski space-time with the shock
wave~\cite{veneziano, thooft}, these coordinates are also closely related to the geodesics. This is
a reason for us to study geodesics in the dS space-time with the shock wave (see Appendix~B for
details and references).

In these coordinates the trapped surface that we seek has two parts, which
are denoted here by $\mathcal{S}_1$ and $\mathcal{S}_2$ in the
respective regions $\Sigma<0$ and $W<0$. They are defined in terms of the
two functions $\Psi_1(\Y,\BY)$ and $\Psi_2(\Y,\BY)$ by
\be
\mathcal{S}_1:\begin{cases}W=0,\\\Sigma=-\Psi_1(\Y,\BY),\end{cases}\qquad
\mathcal{S}_2:\begin{cases}\Sigma=0,\\W=-\Psi_2(\Y,\BY),\end{cases}
\label{ctp}
\ee
with the additional boundary conditions at the shock wave intersection
$\mathcal{C}\subset\{W=\Sigma=0\}$
\be
\Psi_1(\Y,\BY)\Big|_{\mathcal{C}}=0,\qquad
\Psi_2(\Y,\BY)\Big|_{\mathcal{C}}=0,
\label{bc1}
\ee
and
\be
\pp\Psi_1\pb\Psi_2\Big|_{\mathcal{C}}=1.
\label{bc2}
\ee
Because $\mathcal{S}_1$ and $\mathcal{S}_2$ are in the respective
regions $W<0$ and $\Sigma<0$, we also have $\Psi_1(\Y,\BY)>0$ and
$\Psi_2(\Y,\BY)>0$.

The two functions $\Psi_1(\Y,\BY)$ and $\Psi_2(\Y,\BY)$ must be
determined by imposing the condition that the surface they define is
marginally trapped~\cite{HE,poisson,Wald}, i.e., that the congruence of
outgoing null geodesics orthogonal to the surface has zero expansion.

The zero convergence equation for the $D{=}4$ case reduces to the equation
(see Appendix~D)
\be
\label{phi12-eq}
\left(\triangle_{\mathbb{S}^2}+\frac{2}{a^2}\right)\phi_{1,2}(\Y,\BY)=0,
\ee
where
\be
\triangle_{\mathbb{S}^2}=
2\left(1+\frac{\Y\BY}{2a^2}\right)^2\pp\pb
\ee
is the Laplace--Beltrami operator in complex coordinates (see Appendix~D)
and the functions $\phi_{1,2}$ are related to the functions $\Psi_{1,2}$
defining the two halves of the trapped surface and the shock wave shape
functions $H_{1,2}$ (cf.~similar equations in the AdS
case~\cite{gpy,AlvarezGaume:2008fx})
\be
\phi_{1,2}=\frac{2\Psi_{1,2}-H_{1,2}}{1+\Y\BY/2a^2}.
\ee

The next question to be discussed is whether a black hole forms as a result
of the head-on collision of two waves of the type described above. Head-on
collisions preserve rotational symmetry around the axis of motion of
massless particles, the $O(2)$-symmetry in $D=4$. Because a head-on
collision is $O(2)$-symmetric, the functions $\Psi_1(\Y,\BY)$ and
$\Psi_2(\Y,\BY)$ describing the trapped surface are identical and depend
on only the parameter $\rho^2=\Y\BY$: $\Psi_1(\Y,\BY)=\Psi_2(\Y,\BY)=
\Psi(\rho^2)$.

\section{Marginally trapped surface in dS}

\subsection{Solution of the marginally trapped surface equation in dS$_4$}
In the case $\phi_1(\Y,\BY)=\phi_2(\Y,\BY)=\phi(\rho^2)$,
equation~\eqref{phi12-eq} is transformed into the ordinary differential
equation
\be
\label{pho-eq}
\left(1+\frac{\rho^2}{2a^2}\right)^2\left(
\frac{\p^2\phi}{\p\rho^2}+\frac{1}{\rho}\frac{\p\phi}{\p\rho}\right)+
\frac{4\phi}{a^2}=0.
\ee
The solutions of this equation are
\be
\phi=\frac{A(\rho^2-2a^2)+B\bigl((\rho^2-2a^2)\ln\rho+
4a^2\bigr)}{\rho^2+2a^2},
\ee
where $A$ and $B$ are constants. Because we are interested in a solution of
a homogenous equation without singularities, we take $B=0$ and obtain the
expression for the trapped surface function in terms of the shape function
$H$,
\be
\label{Psi}
\Psi=\frac{1}{2}H(\rho)+\left(1+\frac{\rho^2}{2a^2}\right)
\frac{A(\rho^2-2a^2)}{\rho^2+2a^2},
\ee
and also in terms of the shape function $F$,
\be
\label{Psi-F}\Psi=\frac{1}{4}\left(1+\frac{\rho}{2a^2}\right)F(\rho)+
\frac{1}{2a^2}A(\rho^2-2a^2)
\ee
or, more explicitly,
\be
\Psi=\sqrt{2}p\left(1+\frac{\rho^2}{2a^2}\right)
\left(-2+\frac{2a^2-\rho^2}{2a^2+\rho^2}
\ln\left(\frac{2a^2}{\rho^2}\right)\right)+
\frac{1}{2a^2}A(\rho^2-2a^2).
\ee

As mentioned above, the function $\Psi$ must satisfy the following boundary
conditions in the head-on case:
\bea
\Psi\Big|_{\cal C}&=&0,
\\[2mm]
\pp\Psi\pb\Psi\Big|_{\cal C}&=&1.
\eea
It is obvious that the bound $\cal C$ is a circle $\rho=\rho_0=
\mathrm{const}$. We hence have a system of two equations for the two
constants $A$ and $\rho_0$:
\bea
\label{Dirichle}
\sqrt{2}p\left(1+\frac{\rho_0^2}{2a^2}\right)\left(-2+\frac{2a^2-\rho_0^2}
{2a^2+\rho_0^2}\ln\left(\frac{2a^2}{\rho_0^2}\right)\right)+
\frac{1}{2a^2}A\cdot(\rho_0^2-2a^2)=0,
\\[2mm]
\label{rho}
\frac{1}{4a^4\rho_0^2}\left(2\sqrt{2}pa^2+\rho_0^2\left(\sqrt{2}p-A\right)+
\sqrt{2}p\rho_0^2\ln\left(\frac{2a^2}{\rho_0^2}\right)\right)^2=1.
\eea
Substituting $A$ from~\eqref{Dirichle} in expression~\eqref{Psi}, we obtain
\be
\label{Psi-final-dS}
\Psi(\rho)=\sqrt{2}p\left(4\frac{\rho^2-\rho_0^2}{\rho_0^2-2a^2}+
\left(1-\frac{\rho^2}{2a^2}\right)
\ln\left(\frac{\rho_0^2}{\rho^2}\right)\right).
\ee

Equation~\eqref{rho} defining $\rho$ can be also written in terms of the
initial function $F$ defining the shock wave in the plane coordinates,
\be\label{eq-rho-ab}\frac14
\left(1+\frac{\rho_0^2}{2a^2}\right)F'(\rho_0)+
\frac{\rho_0}{2a^2-\rho_0^2}F(\rho_0)+2=0.
\ee
Indeed, normalization condition~\eqref{bc2} can be written in the form
\be
\left(\frac{d\Psi}{d\rho}\right)^2=4\quad\text{or}\quad
\frac{d\Psi}{d\rho}\equiv\Psi'(\rho)=\pm2.
\ee
We choose the minus sign. Because of relation~\eqref{Psi-F} on $\cal C$, we
have
\be
\Psi'(\rho)\Big|_{\cal C}=\frac14\left(1+\frac{\rho_0^2}{2a^2}\right)
F'(\rho_0)+\frac{\rho_0}{2a^2-\rho_0^2}F(\rho_0),
\ee
and we obtain~\eqref{bc2} in the form
\be
\label{eq-rho-m}
\frac14\left(1+\frac{\rho_0^2}{2a^2}\right)
F'(\rho_0)+\frac{\rho_0}{2a^2-\rho_0^2}F(\rho_0)+2=0.
\ee
This equation is universal for an arbitrary dimension $D$. To connect with
the AdS case, this equation can be rewritten in terms of the chordal
coordinate related to $\rho$ by
\be
\rho=a\sqrt{\frac{2}{1-q}}.
\label{q-rho-dS}
\ee
After this change of variable, equation~\eqref{eq-rho-ab} becomes
\be
\label{eq-q-dS}
F'(q_0)+\frac{2}{1-2q_0}F(q_0)+\frac{8a}{\sqrt{2q_0(1-q_0)}}=0.
\ee

We introduce the dimensionless parameter $x_0=\rho_0/a$.
Equation~\eqref{rho} becomes
\be
\label{eq-rho}
f(x_0)=\sqrt{2}\frac{a}{p},
\ee
where
\be
\label{function}
f(x)\equiv\frac{1}{x}\frac{(2+x^2)^2}{2-x^2}.
\ee

We note that in the region $0<x<\sqrt{2}$, the function $f(x)$ has the
positive minimum
\be
f'(x)|_{x=x_{min}}=0,\qquad x_{min}=2-\sqrt{2}.
\ee
Hence, in the case
\be
\label{eq:eta-critical}
\eta\equiv\frac{a}{p}<\frac{1}{\sqrt{2}}\cdot f(x_{min})=4,
\ee
there are no solutions of~\eqref{eq-rho} (see Fig.~1). Therefore, no
solution of the trapped surface equation can be found below the critical
value $\eta_c=4$ of the parameter $\eta$. In terms of the rescaling energy
$\bar{p}=p/G$, where $G$ is the gravitational constant, this relation
gives~\eqref{4cr}.

It is reasonable to consider two limit cases: where $x_0\ll1$ and $p<a$
(when the particle energy is low and/or the spacetime is weakly curved; it
is natural to call this the low-energy limit) and where $x_0$ and $\eta$ are
equal to the critical values (we call this the high-energy critical limit).

From equation~\eqref{eq-rho} in the region $x_0\ll1$, we obtain
\be
x_0\approx\frac{\sqrt{2}p}{a}\quad\text{and}\quad\rho_0\approx\sqrt{2}p.
\ee
And in the critical limit, we obtain
\be
p=\frac{a}{4},\qquad\rho_0=(2-\sqrt{2})a=(8-4\sqrt{2})p.
\ee

\begin{figure}[h!]
\centering
\includegraphics[width=8cm]{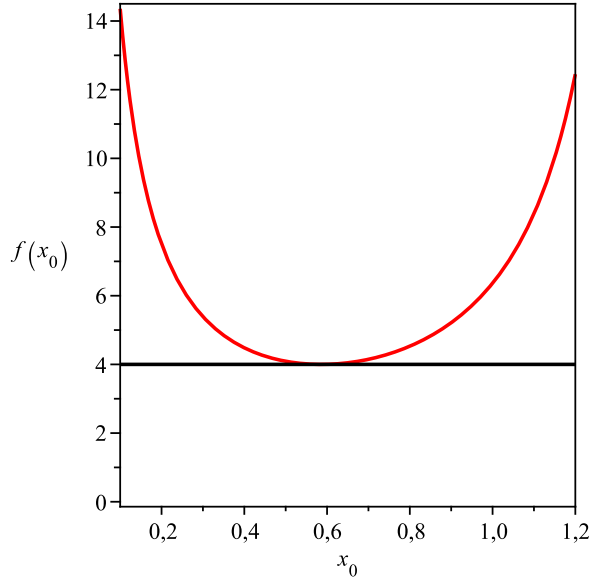}
\caption{The critical value of the parameter $\eta=a/p$ corresponds to the
minimum value of the function $f$ (red line).}
\label{area}
\end{figure}

\subsection{Higher-dimension cases}
Similar calculations can be performed in higher dimensions.
Equation~\eqref{eq-rho-m} has a universal form and is independent of the
dimension $D$. The variable $\rho$ for an arbitrary dimension is related to
$\xi\equiv Z_D/a$ as
\be
\xi=\frac{1-\rho^2/2a^2}{1+\rho^2/2a^2}.
\ee

As examples, we consider the cases $D=5$ and $D=6$. The shape functions have
the forms
\bea
F_5(\xi)&=&\frac{3\sqrt{2}\pi p_5}{a}\frac{2\xi^2-1}{\sqrt{1-\xi^2}}
\\[2mm]
F_6(\xi)&=&\frac{8\sqrt{2}p_6}{a^2}\left(\xi
\ln\left(\frac{1+\xi}{1-\xi}\right)+\frac{2(3\xi^2 -2)}{3(1-\xi^2)}\right),
\eea
where $\xi=Z_D/a$ and $p_D=\bar{p}G_D$. For these cases,
equation~\eqref{eq-rho-m} reduces to
\be
\label{x-0}
f_D(x_0)=C_D\frac{a^{D-3}}{p_D},
\ee
where $C_D$ is a $D$-dependent constant (in particular, $C_3=2/\pi$,
$C_5=32/3\pi$, and $C_6=6\sqrt{2}$) and the functions $f_D(x)$ are given by
\be
f_D(x)=\frac{(2+x^2)^{D-2}}{x^{D-3}(2-x^2)}.
\ee
These functions have positive minima at the points
\bea
x_{min,D}&=&\frac{\sqrt{2(-1+D-2\sqrt{D-2})}}{\sqrt{D-3}},\quad D>3,
\\[2mm]
x_{min,3}&=&0,\quad D=3,
\eea
and the value of the functions $f_D(x)$ at these points,
$f_{0,D}\equiv f_D(x_{min,D})$, give the critical values of the parameter
$\bar{p}$ below which trapped surfaces can be formed,
\bea
\bar{p}&<&\bar{p}_{cr,D},
\label{p-restr}
\\[2mm]
\bar{p}_{cr,D}&=&\frac{a^{D-3}}{G_D}\frac{C_D}{f_{0,D}}.
\label{p-cr}
\eea
We note that this critical effect does not necessarily exclude the trapped
surface in the region of interacting shock waves $W>0$, $\Sigma>0$.

We can also interpret formula~\eqref{p-cr} to mean that for a fixed value of
$\bar{p}$, there is a critical value of the de Sitter radius $a_{cr}$,
\be
a_{cr}\equiv\left(\bar{p}G_D\frac{f_{0,D}}{C_D}\right)^{1/D-3},\quad D>3,
\ee
only above which can trapped surfaces be formed. For $D=3$, the critical
energy is independent of $a$.

In particular, for $D=5$ and $D=6$, we have
\bea
x_{min,5}&=&-1+\sqrt{3},
\\[2mm]
x_{min,6}&=&\frac{\sqrt{6}}{3}
\eea
and
\bea
f_5(x_{5,min})&=&12\sqrt{3},
\\[2mm]
f_6(x_{6,min})&=&\sqrt{6}\frac{2^8}{3^2}.
\eea
Hence, the boundary problem can be solved in $D=5$ and $D=6$ only if the
conditions
\bea
\frac{a^2}{p_5}&>&\frac98\pi\sqrt{3},
\\[2mm]
\frac{a^3}{p_6}&>&\frac{128}{27}\sqrt{3}
\eea
are satisfied. We thus have the same critical effect as in the
four-dimensional dS space-time. The trapped surface can be formed only if
the energy of colliding particles is not very high. Assuming that $x_0\ll1$,
we obtain
\be
\rho_0\approx2\left(\frac{\bar{p}G_D}{C_D}\right)^{1/(D-3)}
\ee
from~\eqref{x-0}.

\subsection{Area of the marginally trapped surface below the critical point}
Knowing $\rho_0$, we can calculate the area of the trapped surface:
\be
{\cal A}_{\mathrm{trap}}=
2\int\limits_{\rho<\rho_0}\sqrt{\det g_{\alpha\beta}}dV,
\ee
where $g_{\alpha\beta}$ is the induced metric on the trapped surface and
$dV$ is an elementary volume in the $(D{-}2)$-dimensional flat space.
Because the corresponding induced metric in the four- and five-dimensional
space-times has the form (the form is similar in higher dimensions)
\bea
g_{\alpha\beta}=\frac1{\N}\begin{pmatrix}\phm0&\phm1\phm\\[1mm]
\phm1&\phm0\phm\end{pmatrix},\qquad
g_{\alpha\beta}=\frac1{\N}\begin{pmatrix}\phm0&\phm1&\phm0\phm\\[1mm]
\phm1&\phm0&\phm0\phm\\[1mm]
\phm0&\phm0&\phm1\phm\end{pmatrix},
\eea
the explicit expression for the area of the trapped surface is
\be
{\cal A}_{\mathrm{trap}}=2\cdot\operatorname{Vol}S^{D-3}
\int\limits_0^{\rho_0}\frac{2^{\frac{D-2}{2}}\rho^{D-3}}{\N^{(D-2)/2}}d\rho=
2\cdot\operatorname{Vol}S^{D-3}\int\limits_0^{\rho_0}
\frac{2^{\frac{D-2}{2}}\rho^{D-3}}{(1+\rho^2/2a^2)^{D-2}}d\rho.
\ee
In particular,
\bea
{\cal A}^4&=&8\pi\frac{a^2\rho_0^2}{2a^2+\rho_0^2},
\\[2mm]
{\cal A}^5&=&4\sqrt{2}\pi a^3\left(\sqrt{2}
\arctan\left(\frac{\sqrt{2}}{2}\frac{\rho_0}{a}\right)+
\frac{2a\rho_0(\rho_0^2-2a^2)}{(2a^2+\rho_0^2)^2}\right).
\eea

We have
\bea
{\cal A}^4_{\mathrm{LE}}&\approx&8\pi\frac{a^2p^2}{a^2+p^2}\approx
8\pi p^2\left( 1-\frac{p^2}{a^2}\right),
\\[2mm]
{\cal A}^5_{L\mathrm{E}}&\approx&\frac{16\sqrt{2}\pi}{3}\rho_0^3\approx
\sqrt{\frac{3}{2\pi}p_5^3}
\eea
in the low-energy limit and
\bea
{\cal A}^4_{C\mathrm{r}}&=&(4-2\sqrt{2})\pi a^2,
\\[2mm]
{\cal A}^5_{\mathrm{Cr}}&=&\frac{8\sqrt{2}\pi}{(3-\sqrt{3})^2}(\sqrt{3}-2)
\left(1-3\sqrt{2}\arctan\frac{\sqrt{2}(\sqrt{3}-1)}{2}\right)a^3
\eea
at the critical point.

\section{Concluding Remarks}

We have studied the formation of marginally trapped surfaces in head-on
collisions of two shock waves in the dS space-time for $D\le3$. For $D\le4$,
we found the critical value of of the shock wave energy dependent on the dS
radius, above which the trapped surface equation has no solution. This
critical behavior is similar to that found in~\cite{AlvarezGaume:2008fx}
and is also reminiscent of the behavior encountered in numerical
simulations of gravitational collapse~\cite{choptuikPRL,reviews}. For $D=3$,
the critical energy above which there is no trapped surface is independent
of the cosmological constant.

$$~$$

\section*{Acknowledgments}

I.\,A. is grateful to I.~Volovich for the fruitful discussions. We are
supported in part by the RFBR grant 08-01-00798 and 09-01-12179 and by the Federal Agency of Science
and Innovations (contract ¹ 02.740.11.5057).

\newpage
\section{Appendices}

\appendix
\section{Geometric view of shock waves in plane coordinates}

\begin{figure}[h!]
\centering
\includegraphics[width=5cm]{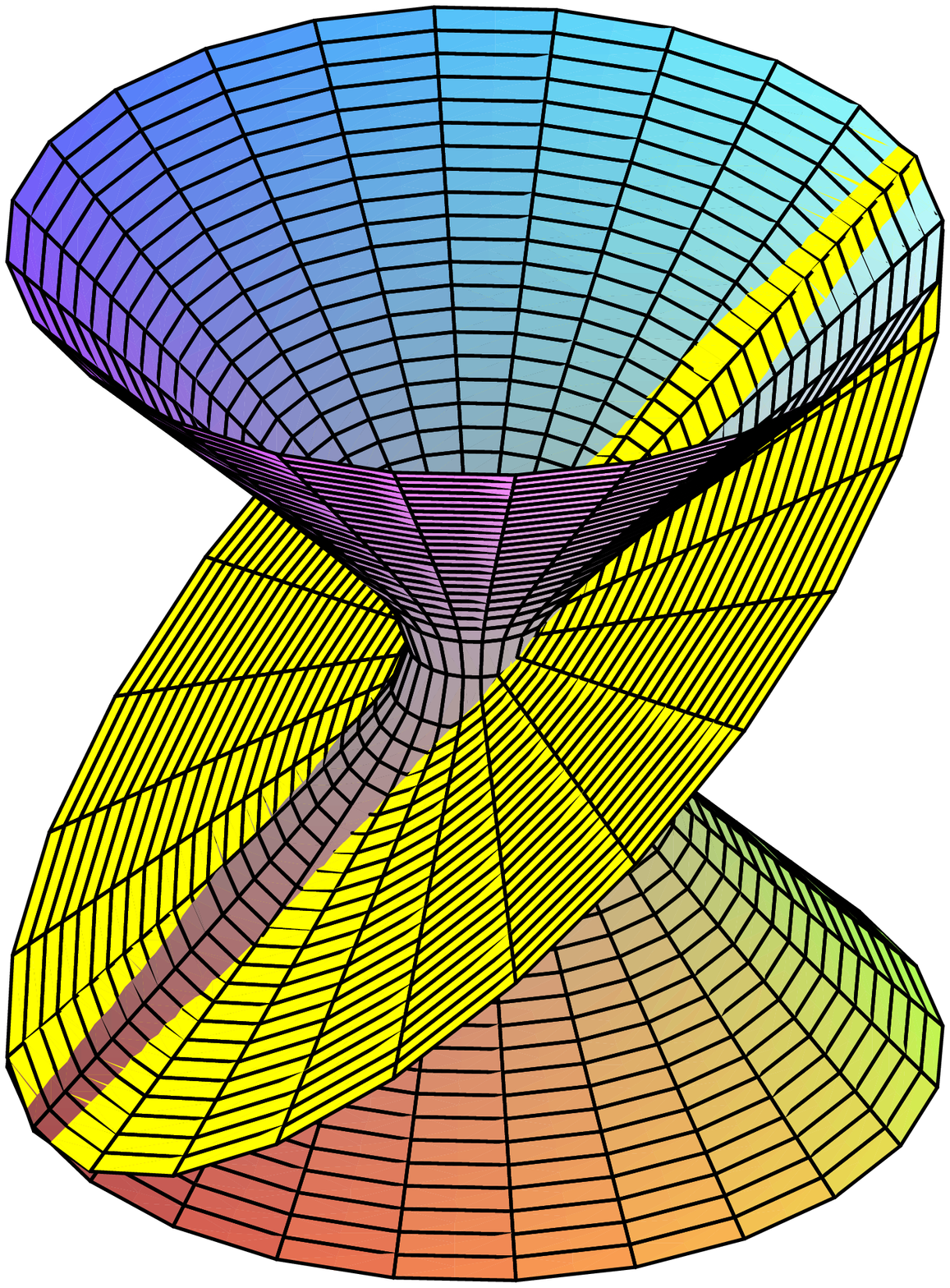}A
\includegraphics[width=5cm]{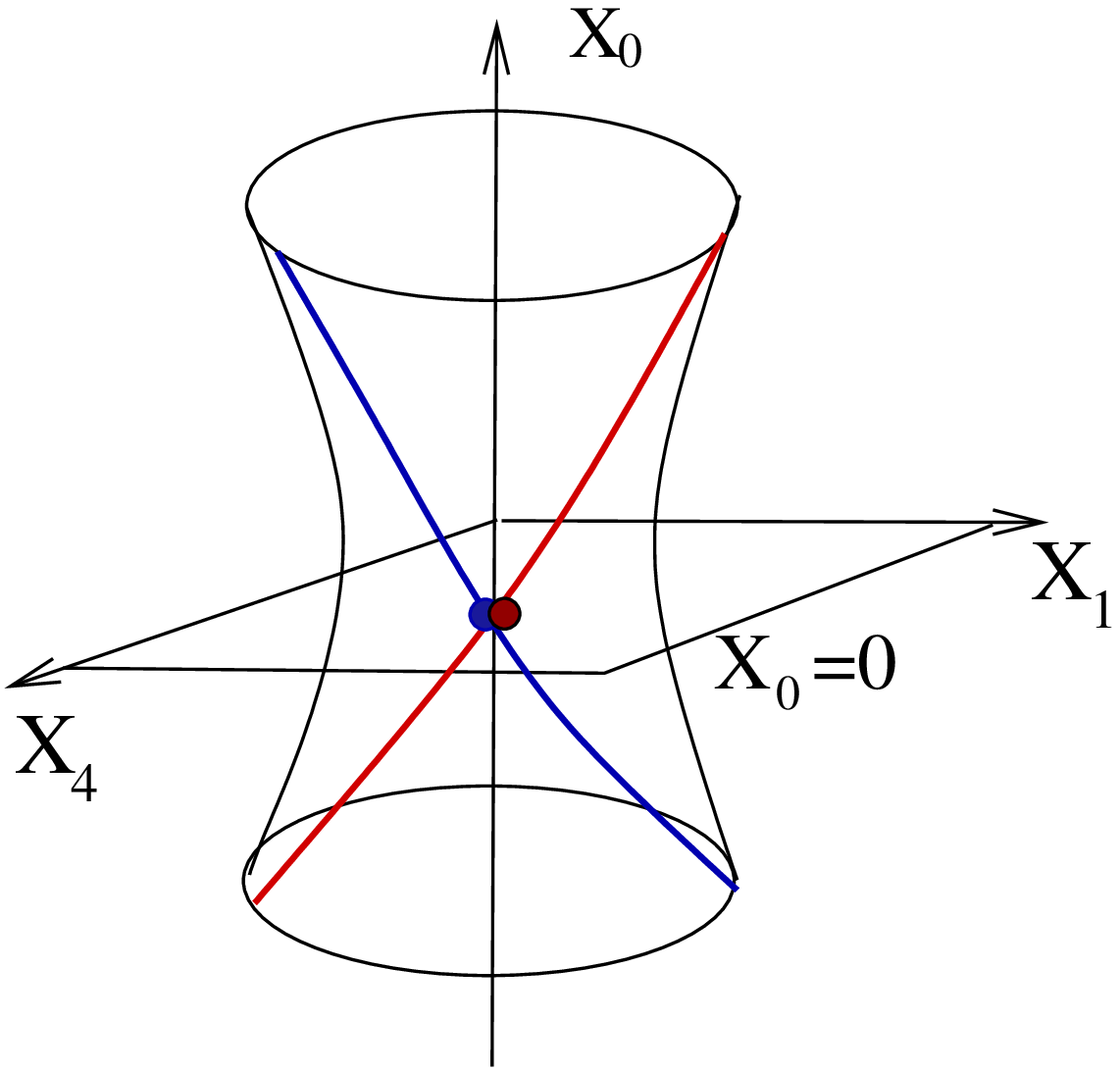}B
\caption{A.~A single shock wave in the dS space can be represented as the
intersection of the hyperboloid and the plane $x^0-x^1=0$ ($x^2$ and $x^3$
are suppressed). B.~Two shock waves in the dS space. A collision of two
shock waves occurs at $x^0=0$ and corresponds to the collision of red and
yellow balls.}
\label{Oneshock}
\end{figure}

A single shock wave in the $D$-dimensional dS space is shown in
Fig.~\ref{Oneshock}.A. The dS space is represented as a hyperboloid embedded
into the $(D{+}1)$-dimensional Minkowski space-time. The presented shock
wave is located on the intersection of the hyperboloid and the plane
$x^0-x^1=0$,
$$
u=\frac{x^0+x^1}{\sqrt2},\qquad v=\frac{x^0-x^1}{\sqrt2}.
$$
The coordinates $x_2$ and $x_3$ are suppressed in this figure.

Two shock waves colliding at $u=v=0$ are shown in Fig.~\ref{Oneshock}.B.

\begin{figure}[h!]
\centering
\includegraphics[width=3.5cm]{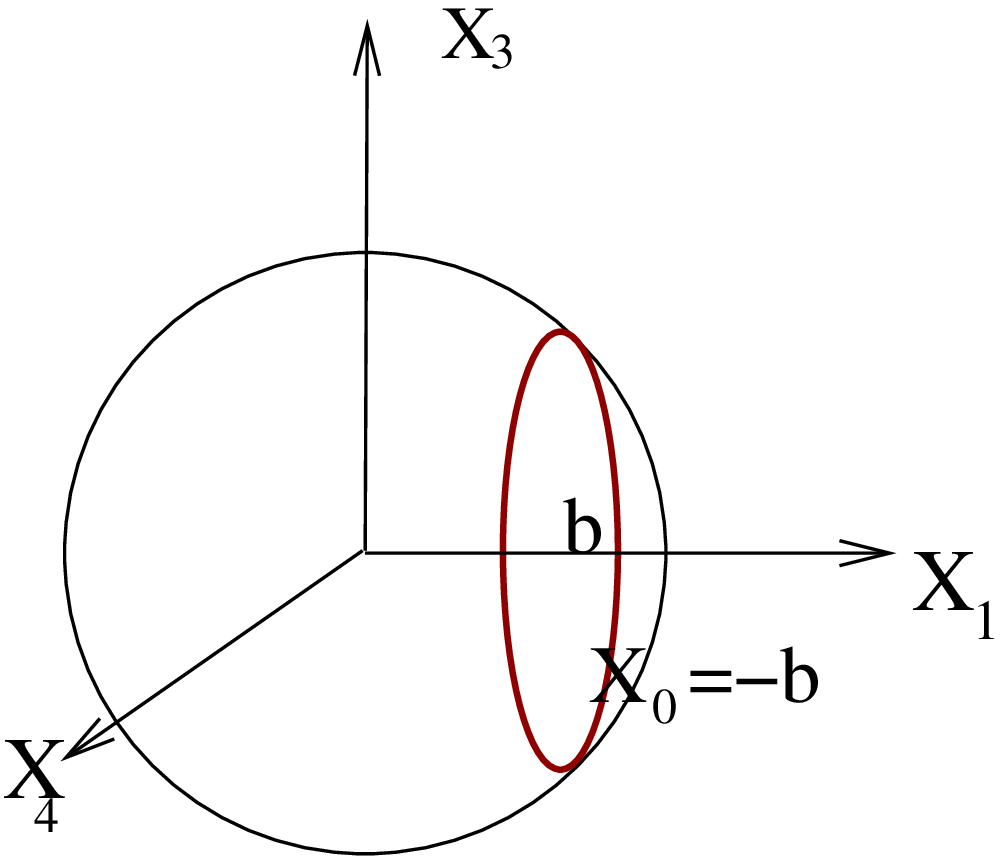}$\qquad$
\includegraphics[width=3.5cm]{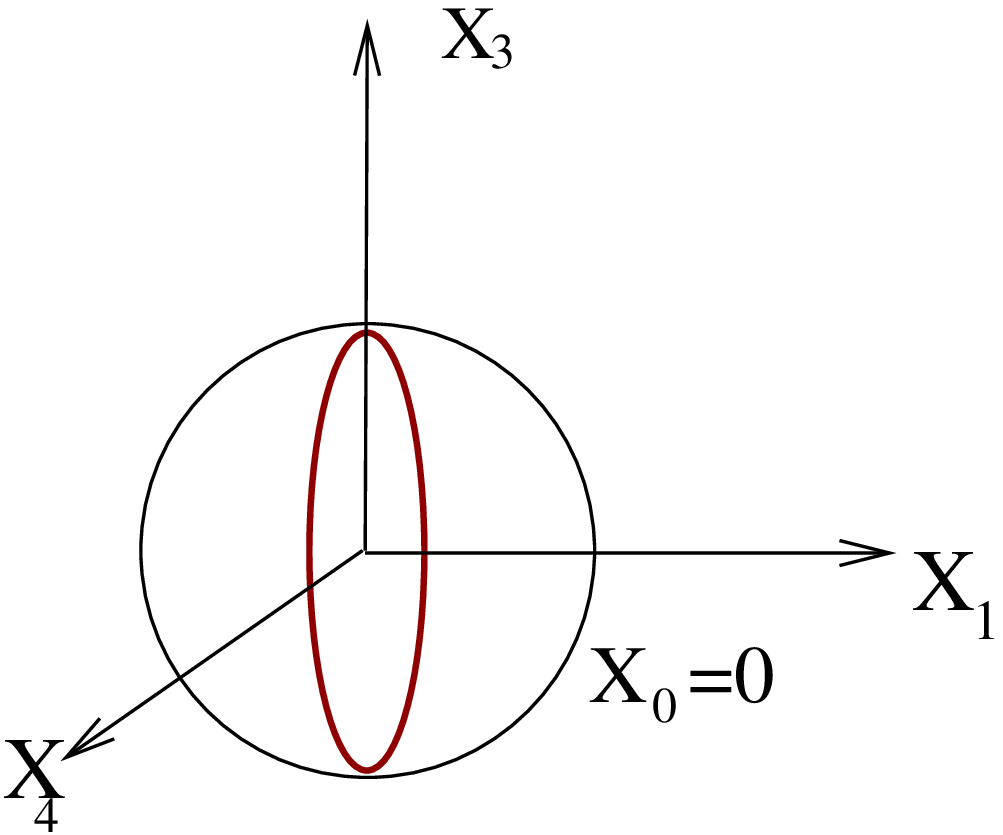}$\qquad$
\includegraphics[width=3.5cm]{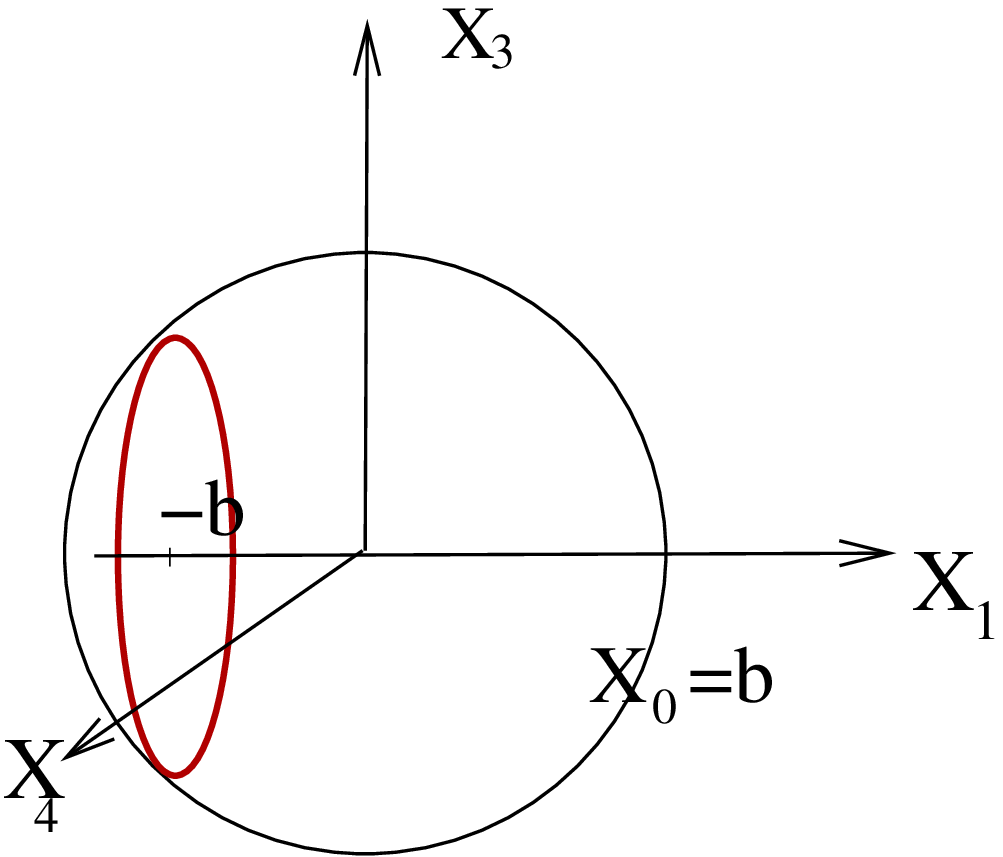}
\caption{A shock wave in the dS space at different instants of
``$x^0$ time.''}
\label{Twoshock}
\end{figure}

We can also make an animation and draw the position of the single shock
wave at discrete instants. In Fig.~\ref{Twoshock}, we see that the shock
wave is a nonexpanding one.

\newpage

\section{Solution of the geodesics equation}

\subsection{The $\sigma$-model ($n$-field) approach}
\par
In this section, we derive and solve the null-geodesic equations in the dS
space-time with a nonexpanding shock-wave. We note that applying the
embedding theorems~\cite{Podolsky:2001vu} requires a delicate analysis
because of the nonsmoothness of the metric with the shock wave.

We use an analytical approach similar to the $\sigma$-model ($n$-field)
approach and start from the Lagrangian
\be
{\cal L}=\int d\tau\left[\frac{dx^M(\tau)}{d\tau}G_{MN}\bigl(x(\tau)\bigr)
\frac{dx^N(\tau)}{d\tau}-
\lambda\left(x^M(\tau)g_{MN}x^N(\tau)-a^2\right)\right],
\ee
where $M,N=0,\dots,D$, $g_{MN}$ is the metric defined the hyperboloid, and
$G_{MN}(x(\tau))$ is the metric deformed by the shock wave,
\be
G_{MN}=g_{MN}+F\delta(u)du^2.
\ee
The Euler equations for this Lagrangian are
\bea
G_{MN}\frac{d^2x^N(\tau)}{d\tau^2}+
G_{MN}\Gamma^N_{KL}\frac{dx^K(\tau)}{d\tau}\frac{dx^L(\tau)}{d\tau}+
\lambda g_{MN}x^N(\tau)&=&0,
\label{eq:euler-eq}
\\[2mm]
x^M(\tau)g_{MN}x^N(\tau)-a^2&=&0.
\label{eq:bound}
\eea
Using $u\delta(u)=0$, we obtain
\be
G^{MN}g_{NK}x^K=x^M,
\ee
and equations~\eqref{eq:euler-eq} and~\eqref{eq:bound} reduce to
\be
\label{eq:F-EOM}
\ddot{x}^N+\Gamma^N_{KL}\dot x^K\dot x^L+\lambda\,x^N=0,\qquad
\lambda=-\frac{1}{a^2}(x,\ddot{x}+\Gamma _{KL}\dot x^K\dot x^L\,)_G.
\ee
Here and hereafter, $(x,y)_G=G_{MN}x^My^N$. The nonvanishing components of
the connection $\Gamma^M_{NK}$ are
\be
\label{Gamma}
\Gamma^v_{uu}=-\frac12F\delta'(u),\qquad
\Gamma^v_{ui}=-\frac12F_{,i}\delta(u),\qquad
\Gamma^i_{uu}=-\frac12F_{,i}\delta(u).
\ee
Taking $(x,x)_g=a^2$ and $G_{\mu\nu}\dot x^\mu\dot x^\nu=0$ into account, we
obtain
\be
\label{exp-lambda-resul}
\lambda=\frac1{2a^2}(-F+x^iF_{,i})\delta(u)\dot u^2.
\ee
Substituting this expression in~\eqref{eq:F-EOM} and taking~\eqref{Gamma}
into account, we obtain
\bea
\label{ddotU}
\ddot u&=&-\frac1{2a^2}(-F+x^iF_{,i})\delta(u)\dot u^2u,
\\[2mm]
\ddot v-\frac1F\delta^\prime(u)\dot u^2-F_{,i}\delta(u)\dot u\dot x^i&=&
-\frac1{2a^2}(-F+x^iF_{,i})\delta(u)\dot u^2v,
\\[2mm]
\ddot x^i-\frac12F_{,i}\delta(u)\dot u^2&=&
-\frac1{2a^2}(-F+x^iF_{,i})\delta(u)\dot u^2x^i.
\eea
Noting that the right-hand side of~\eqref{ddotU} vanishes, we obtain
$\ddot u=0$. Taking $\tau=u$, we obtain
\bea
\label{GV-dsSW}
\ddot v-\frac12F\delta'(u)-F_{,i}\delta(u)\dot x^i&=&
-\frac1{2a^2}(-F+x^iF_{,i})v\delta(u),
\\[2mm]
\label{GX-dsSW}
\ddot x^i-\frac12F_{,i}\delta(u)&=&-\frac1{2a^2}(-F+x^jF_{,j})x^i\delta(u).
\eea
We now use the ansatz (by analogy with the case of the shock wave in the
Minkowski space-time~\cite{thooft,veneziano})
\bea
\label{tot-1-ds}
v&=&v_0+v_1u+Q(x^j_0)\theta(u)+R(x^j_0)\theta(u)u,
\\[2mm]
\label{tot-2-ds}
x^i&=&x_{i0}+x_{i1}u+x_{if}\theta(u)+S_i(x^j_0)\theta(U)U,
\\[2mm]
\label{tot-1d-ds}
\dot v&=&v_1+Q(x^j_0)\delta(u)+R(x^j_0)\theta(u),
\\[2mm]
\label{tot-2d-ds}
\dot x^i&=&x_{i1}+D_i(x^j_0)\delta(u)+S_i(x^j_0)\theta(u),
\\[2mm]
\label{tot-1dd-ds}
\ddot v&=&Q(x^j_0)\delta^\prime(u)+R(x^j_0)\delta(u),
\\[2mm]
\label{tot-2dd-ds}
\ddot x^i&=&D_i(x^j_0)\delta^\prime(u)+S_i(x^j_0)\delta(u).
\eea
Along the geodesics, we have the identity
\be
\label{delta-prime-gener}
F(x^i)\delta'(u)=F(x^i_0)\delta'(u)-F_{,i}(x^i_0)\dot x^i\delta(u)
\ee
(a similar identity for the flat space-time case was used
in~\cite{veneziano}). Transforming our anzatz in order to cancel
$\delta ^2(u)$, we now obtain
\be
\begin{aligned}
v&=v_0+v_1u+Q(x^j_0)\theta(u)+R(x^j_0)\theta(u)u,
\\[2mm]
x^i&=x^i_0+x^i_1u+S_i\theta(u)u
\end{aligned}
\label{tot-4}
\ee
with the bounds
\bea
x_{i0}^2=a^2,
\\[2mm]
v_0=x_0^ix_1^i,
\\[2mm]
v_1=\frac12x_1^{i2}.
\eea
And hence
\bea
\label{AVf}
Q&=&\frac12F,
\\[2mm]
\label{AVd}
R&=&\frac12F_{,i}x^i_1+\frac1{2a^2}(F-x^i_0 F_{,i})v_0+\frac18 F_{,i}^2+
\frac1{8a^2}(F^2-(x^i_0 F_{,i})^2),
\\[2mm]
\label{AXd}
S_i&=&\frac12F_{,i}+\frac1{2a^2}(F- x^j_0F_{,j})x^i_0.
\eea

We can take $x^i_1=0$ for simplicity. This gives $v_0=v_1=0$, and we have
\be
\begin{aligned}
Q&=\frac12F
\\[2mm]
R&=\frac18F_{,i}^2+\frac1{8a^2}(F^2-(x^i_0 F_{,i})^2),
\\[2mm]
S_i&=\frac12F_{,i}+\frac1{2a^2}(F- x^j_0 F_{,j})x^i_0.
\end{aligned}
\label{eq:rs-simp}
\ee

\subsection{Focusing of geodesics}
\begin{figure}[h!]
\begin{center}
\setlength{\unitlength}{0.8mm}
\begin{picture}(75,60)
\put(0,35){$x^2$}
\put(5,-2){\includegraphics[height=6cm]{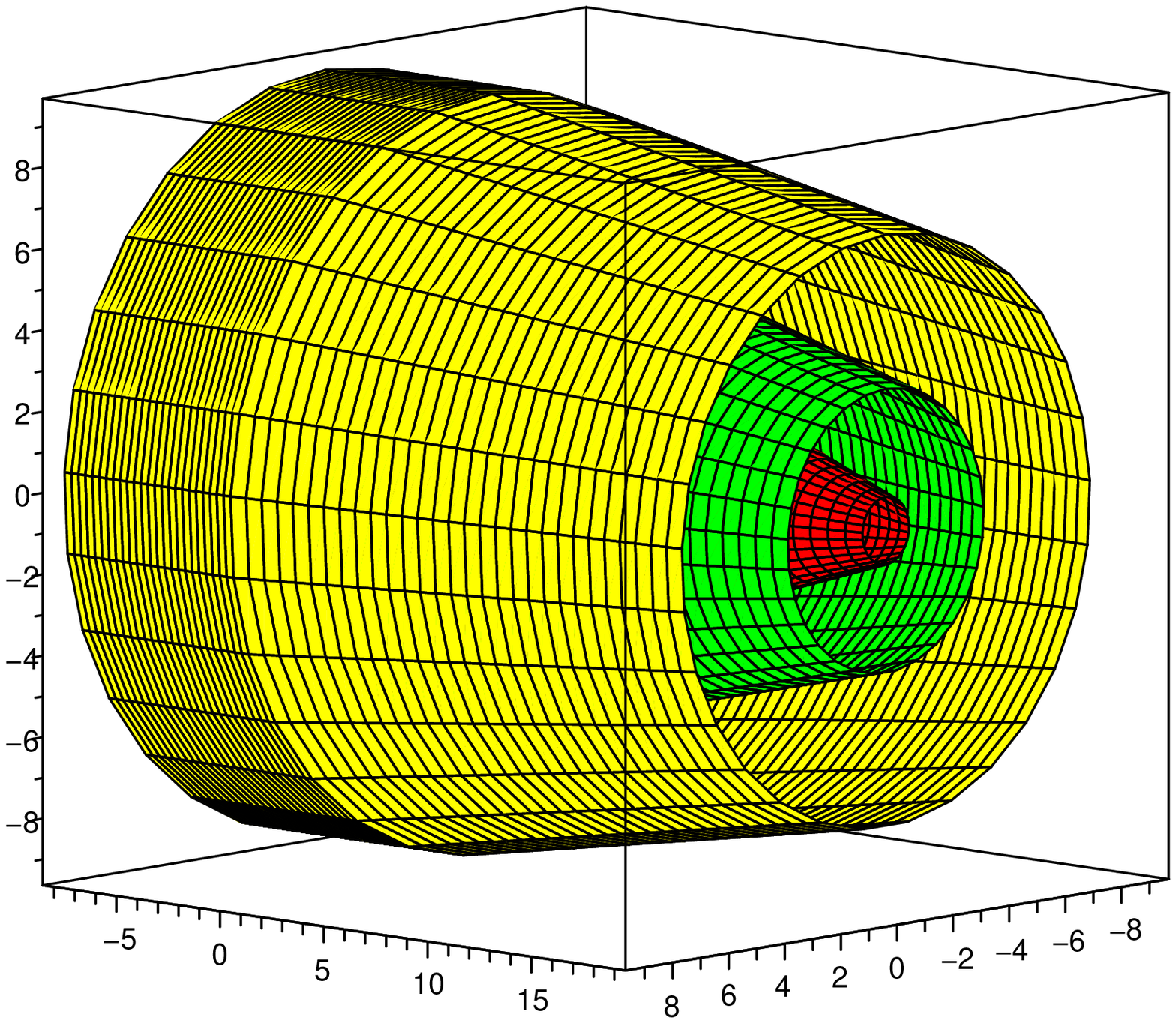}}
\put(20,2.5){$u$}\put(55,1){$x^3$}
\end{picture}$\qquad$
\begin{picture}(75,60)
\put(0,35){$x^2$}
\put(5,-2){\includegraphics[height=6cm]{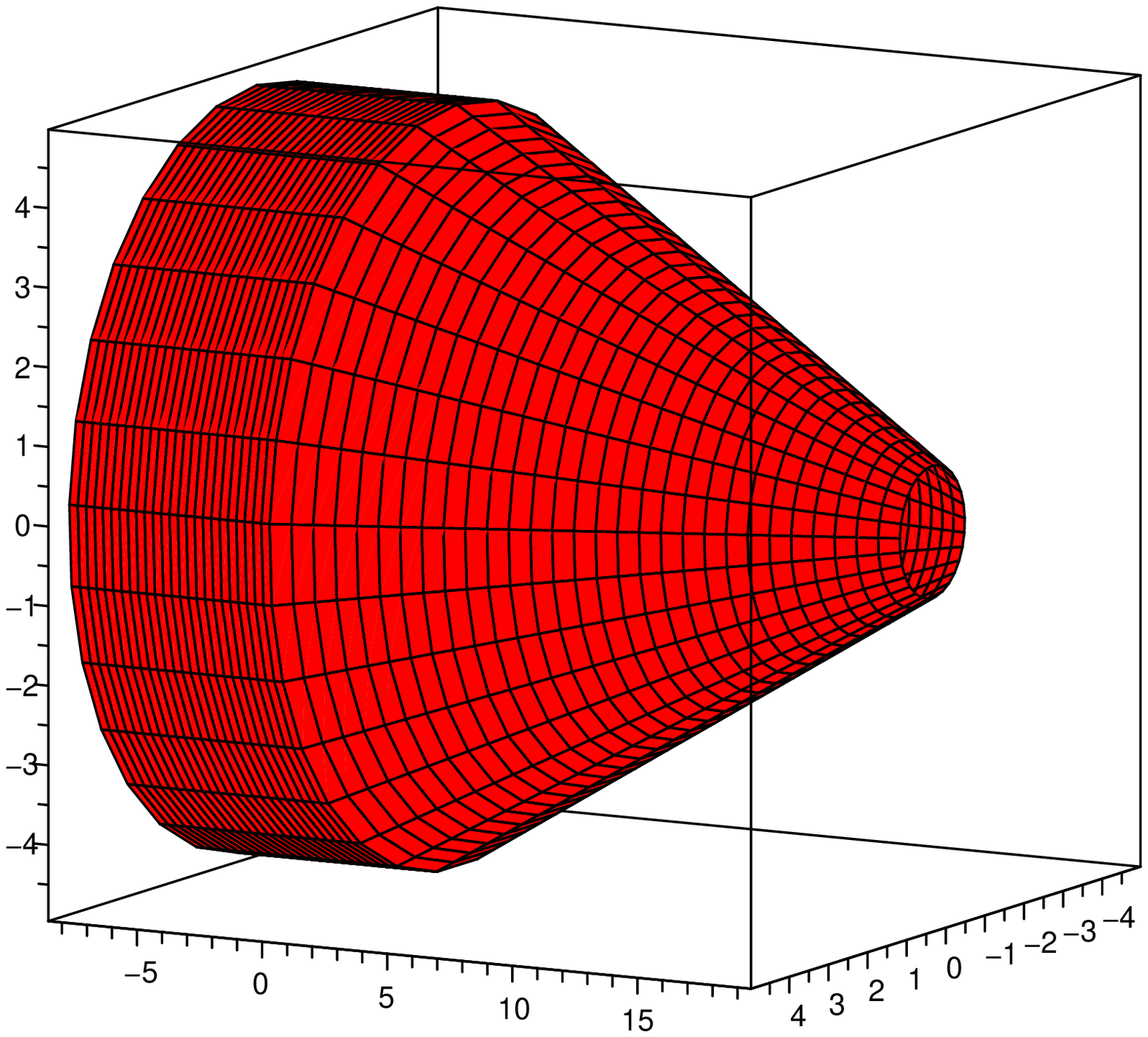}}
\put(25,2.5){$u$}\put(65,1){$x^3$}
\end{picture}
\end{center}
\vspace*{-3mm}
\caption{Focusing of geodesics: in the dependent coordinates, the focal
length changes along with the initial conditions.}
\label{graph}
\end{figure}

Before deriving the explicit expression for the trapped surface, we study
the structure of geodesic beams in terms of the dependent plane coordinates.
We consider points that are initially on the surface
$x_{20}^2+x_{30}^2+x_{40}^2=a^2$ and change to the angular coordinates:
$$
x_{20}=a\cdot\sin\phi\cdot\sin\theta,\qquad x_{30}=
a\cdot\sin\phi\cdot\cos\theta,\qquad x_{40}=a\cdot\cos\phi,
$$
$$
F=4\sqrt{2}p\left(-2+\frac{x_{40}}a
\ln\left(\frac{a+x_{40}}{a-x_{40}}\right)\right)=
4\sqrt{2}p\left(-2+\cos\phi
\ln\left(\frac{1+\cos\phi}{1-\cos\phi}\right)\right).
$$
It is obvious that the term $S_i(x^j_0)\theta(u)u$ in
formula~\eqref{tot-2d-ds} leads to the refraction of the geodesics. The
refraction coefficients are $S_i(x^j_0)$. For the refraction coefficients in
the angular parameterization, we have
\be
S_2=-\sqrt{2}\,\frac pa\frac{\sin\theta}{\sin\phi},\qquad
S_3=-\sqrt{2}\,\frac pa\frac{\cos\theta}{\sin\phi},\qquad
S_4=\frac{\sqrt2}2\frac pa\ln\left(\frac{1+\cos\phi}{1-\cos\phi}\right).
\ee
The coordinates $x_2$ and $x_3$ correspond to the physical coordinates
related to the size of the beam. Hence, we can easily find the value of the
affine parameter $u$, which determines the focal point of the beam (the
point $x_2=x_3=0$):
$$
u=\frac{a^2}{\sqrt{2}p}\cdot\sin^2\phi.
$$
The ``focal length'' is different for each value of the parameter $\phi$,
i.e., for each circular ring of the beam cross section. This statement is
illustrated in Fig.~\ref{graph}. Of course, this only roughly explains why
the trapped surface can exist. To find a physical meaning for this
consideration, we must change coordinates to independent ones.

\vspace*{5mm}

\newpage
\section{The shock wave in the independent coordinates}

\subsection{From plane coordinates to independent coordinates}
To study the structure of the space-time in terms of the independent
four-dimensional coordinates, it is convenient to use the complex conformal
flat coordinates
\be
w=\frac{2au}{x^4+a},\qquad\sigma=\frac{2av}{x^4+a},\qquad
\z=\frac{\sqrt2 a}{x^4+a}(x^2+ix^3).
\label{eq:rchange}
\ee
In these coordinates, the shock wave metric is
\be
ds^2=\frac{-2dw\,d\sigma+2d\z\,d\bz+2H(\z,\bz)\delta(w)\,dw^2}
{[1-(w\sigma-\z\bz)/2a^2]^2},
\label{com-sw}
\ee
where
\be
\label{eq:complex-sw}
H(\z,\bz)=\frac12\left(1+\frac{1}{2a^2}\z\bz\right)
F\left(\frac{1-\z\bz/2a^2}{1+\z\bz/2a^2}\right)
\ee
and F is given by~\eqref{eq:F4}. Hence,
\be
\label{eq:complex-sw-4}
H(\z,\bz)=2\sqrt2p\left(1+\frac{1}{2a^2}\z\bz\right)
\left(-2+\frac{1-\z\bz/2a^2}{1+\z\bz/2a^2}
\ln\left(\frac{2a^2}{\z\bz}\right)\right).
\ee

\subsection{An analogue of the D'Eath and Payne coordinates\\
(from the coordinates $w,\sigma,\z$ to the coordinates $W,\Sigma,\Y$)}
To eliminate $\delta(u)$ from the metric, we use a coordinate change
analogous to the change introduced in~\cite{dp},
\be
\begin{aligned}
w&=W,
\\[2mm]
\sigma&=\Sigma+H(\Y,\BY)\theta(W)+W\theta(W)H_{\Y}H_{\BY},
\\[2mm]
\z&=\Y+W\theta(W)H_{\BY},
\end{aligned}
\label{eq:trans1}
\ee
where $H_{\Y}=\pp H(\Y,\BY)$. In these coordinates, we obtain the metric
\be
ds^2=\frac{-2dW\,d\Sigma+2|d\Y+W\theta(W)(H_{\Y\BY}d\Y+H_{\BY\BY}d\BY)|^2}
{[1-(W\Sigma-\Y\BY+W\theta(W)G)/2a^2]^2},
\label{com-smooth}
\ee
where $G=H-\Y H_\Y-\BY H_{\BY}$ and $H(\Y,\BY)$ depends on $\Y,\BY$ as
$H(\z,\bz)$ given by~\eqref{eq:complex-sw} depends on $\z,\bz$.

\subsection{Geodesics in terms of the independent complex conformal flat
coordinates}
We have the expressions for geodesics in terms of the dependent coordinates.
Using coordinate change~\eqref{eq:rchange}, we obtain an expression for
geodesics in terms of the independent coordinates. In the first order of the
parameter $u$, we obtain
\be
\label{eq:series-w}
\begin{aligned}
w(u)&=w_1u+\dots,
\\[2mm]
\sigma(u)&=\sigma_0+\sigma_1u+\cdots\equiv
\sigma_{0c}+\sigma_{0\theta}\theta(u)+(\sigma_{1c}
+\sigma_{1\theta}\theta(u))u+\dots,
\\[2mm]
\z(u)&=\z_0+\z_1u+\cdots\equiv
\z_{0c}+\z_{0\theta}\theta(u)+(\z_{1c}+\z_{1\theta}\theta(u))u+\dots,
\end{aligned}
\ee
where
\bea
w_1&=&\frac{2}{1+x_0^4/a},
\\[2mm]
\sigma_{0c}&=&\frac{2v_0}{1+x_0^4/a},
\\[2mm]
\sigma_{0\theta}&=&\frac{2Q(x^i_0)}{1+x_0^4/a},
\\[2mm]
\label{sv1}
\sigma_{1c}&=&2\left(\frac{v_1}{1+x_0^4/a}-
\frac{x_1^4}{a}\frac{v_0}{(1+x_0^4/a)^2}\right),
\\[2mm]
\sigma_{1\theta}&=&2\left(\frac{R(x^i_0)}{1+x_0^4/a}-
\frac{Q(x^i_0)x_1^4}{a(1+x_0^4/a)^2}-\frac{QS^4(x^i_0)}{a(1+x_0^4/a)^2}-
\frac{S^4(x_0^4)v_0}{a(1+x_0^4a)^2}\right),
\\[2mm]
\z_{0c}&=&\frac{\sqrt{2}z_0}{1+x_0^4/a},
\\[2mm]
\z_{0\theta}&=&0,
\\[2mm]
\label{zeta1c}
\z_{1c}&=&\frac{\sqrt{2}z_1}{1+x_0^4/a}-
\frac{x_1^4}{a}\frac{\sqrt{2}z_0}{(1+x_0^4/a)^2},
\\[2mm]
\z_{1\theta}&=&\frac{\sqrt{2}{\cal S}}{1+x_0^4/a}-
\frac{S^4}{a}\frac{\sqrt{2}z_0}{(1+x_0^4/a)^2},
\eea
where the complex variables ${\cal S},\,z_0,\,z_1$ are related to
$S^i,\,x_0^i,\,x_1^i$ by
\bea
{\cal S}&=&S^2+iS^3,
\\[2mm]
z_0&=&x^2_0+ix^3_0,\qquad z_1=x^2_1+ix^3_1.
\eea
We see that there is a discontinuity only for the $\sigma$ variable.

\subsection{Geodesics in terms of the independent smooth coordinates}
Using relations~\eqref{eq:trans1} between the initial independent
coordinates and the smooth independent coordinates, we obtain the expression
for geodesics in terms of the smooth independent coordinates up to the
second order:
\bea
\Sigma(w)&=&\bigl(\Sigma_{0c}+\Sigma_{0\theta}\theta(w)\bigr)+
\bigl(\Sigma_{1c}+\Sigma_{1\theta}\theta(w)\bigr)w+
\Sigma_2\frac{w^2}{2}+\dots,
\label{Sigma}
\\[2mm]
\Y(w)&=&\Y_0+\bigl(\Y_{1c}+\Y_{1\theta}\theta(w)\bigr)w+
\Y_2\frac{w^2}{2}+\dots,
\\[2mm]
\BY(w)&=&\BY_0+\bigl(\BY_{1c}+\BY_{1\theta}\theta(w)\bigr)w+
\BY_2\frac{w^2}{2}+\dots,
\label{Y-bar}
\eea
where
\bea
\Sigma_{0c}&=&\sigma_0,
\label{S-S0c}
\\[2mm]
\Sigma_{0\theta}&=&0,
\label{S-St}
\\[2mm]
\Sigma_{1c}&=&\sigma_{1c},
\label{S-S1}
\\[2mm]
\Sigma_{1\theta}&=&\sigma _{1\theta}-H_\Y(\Y_0,\BY_0)\Y_1-
H_{\BY}(\Y_0,\BY_0)\BY_1-w_1H_{\Y}(\Y_0,\BY_0)H_{\BY}(\Y_0,\BY_0),\qquad\quad
\\[2mm]
\Y_0&=&\z_{0c},
\\[2mm]
Y_{1c} &=&\z_{1c},
\\[2mm]
\label{Y-z1}
Y_{1\theta}&=&\z_{1\theta}-w_1H_{\BY}(\Y_0,\BY_0).
\eea

\section{Trapped surface equation}
In the case of two shock waves, we deal with
\bea
ds^2&=&2g_{W\Sigma}dWd\Sigma+
g_{\Y\Y}d\Y d\Y+2g_{\Y\BY}d\Y d\BY+g_{\BY\BY}d\BY d\BY=
\nonumber
\\[2mm]
&=&\frac{-2dW\,d\Sigma +2|d\Y+\bigl(\Sigma\theta(\Sigma)+W\theta(W)\bigr)
(H_{\Y\BY}d\Y+H_{\BY\BY}d\BY)|^2}
{[1-(W\Sigma-\Y\BY+(\Sigma\theta(\Sigma)+W\theta(W))G/2a^2)]^2}.
\label{2shocks}
\eea
Because of the notation symmetry ($W\leftrightarrow\Sigma$), we may analyze
only one part of the trapped surface.

The null geodesics passing through the surface
\be
\label{surface}
\Sigma=\Psi(\Y,\BY),\qquad W=0
\ee
can be specified by the tangent vectors (see Subsections C.3 and C.4):
\be
\xi^W=w_1,\qquad\xi^\Sigma=\sigma_1,\qquad
\xi^{\Y}=\z_1,\qquad\xi^{\BY}=\bz_1,
\ee
where
\be
\label{sigma1}
\sigma_1=-\frac1{2g_{W\Sigma}w_1}
(g_{\Y\Y}\z_1^2+2g_{\Y\BY}\z_1\bz_1+g_{\BY\BY}\bz_1^2).
\ee
We also suppose that
\be
(\xi,K_a)=0,\qquad a=\z,\bz,
\ee
for
\be
K_a^M=(0,-\p_a\Psi,\delta_a^b)
\ee
(our calculations are very close to those in the review section
in~\cite{Nastase:2005rp}). This gives
\bea
\label{K-Psi-m}
-g_{W\Sigma}\p_a\Psi w_1+g_{ab}\z^b=0.
\eea
From~\eqref{K-Psi-m}, we have
\be
\label{zeta-i}
\z^a=w_1g^{ab}g_{W\Sigma}\p_b\Psi,
\ee
where $g^{ab}$ is the inverse metric for $g_{ab}$. Therefore, we have
\bea
\xi^W&=&w_1,
\\[2mm]
\xi^\Sigma&=&\sigma_1=-\frac{w_1g_{W\Sigma}}{2\det(g^{ab})}(g_{\BY\BY}\pb
\Psi\pb\Psi-2g_{\Y\BY}\pp\Psi\pb\Psi+g_{\Y\Y}\pb\Psi\pb\Psi),\qquad\quad
\\[2mm]
\xi^{\Y}&=&\z_1=\frac{w_1g_{W\Sigma}}{\det(g^{ab})}
(g_{\BY\BY}\pp\Psi -g_{\Y\BY}\pb\Psi),
\\[2mm]
\xi^{\BY}&=&\bz_1=\frac{w_1g_{W\Sigma}}{\det(g^{ab})}
(g_{\Y\Y}\pb\Psi-g_{\Y\BY}\pp\Psi).
\eea
For $\xi_M=g_{MN}\xi^N$ at the point $W=0$, we have
\be
\xi_M=\left(-\frac{\pp\Psi\pb\Psi}{1+\Y\BY/2a^2},-\frac{1}{1+\Y\BY/2a^2},
-\frac{\pp\Psi}{1+\Y\BY/2a^2},-\frac{\pb\Psi}{1+\Y\BY/2a^2}\right).
\ee
For the convergence at the surface $W=0,\Sigma=-\Psi(\z,\bz)$, we obtain
\be
\theta=h^{MN}\nabla _N\xi_M,
\ee
where
\be
h^{MN}=K^M_{\alpha} g^{\alpha\beta}K^N_{\beta}
\ee
and $g^{\alpha\beta}$ is the inverse of the metric induced on the trapped
surface,
\be
g_{\alpha\beta}=K^M_{\alpha}g_{MN}K^N_{\beta}=\begin{pmatrix}
g_{\Y\Y}&g_{\Y\BY}\\g_{\BY\Y}&g_{\BY\BY}\end{pmatrix}.
\ee

The components of the connection for metric~\eqref{2shocks} in the
coordinates $X^M=(W,\Sigma,\Y,\BY)$, $M,N=0,1,2,3$, are
\bea
\Gamma^1_{11}&=&-\frac{\p_\Sigma{\N}}{\N},\qquad
\Gamma^1_{12}=-\frac{\pp\N}{2\N},\qquad
\Gamma^2_{12}=-\frac{\p_\Sigma\N}{2\N},
\\[2mm]
\Gamma^1_{22}&=&\frac{H_{\Y\Y}}{2},\qquad
\Gamma^2_{22}=-\frac{\pp\N}{\N},\qquad\Gamma^1_{13}=-\frac{\pb\N}{2\N},
\\[2mm]
\Gamma^3_{13}&=&-\frac{\p_\Sigma\N}{2\N},\qquad
\Gamma^1_{23}=-\frac{1}{2\N}(\p_W\N -H_{\Y\BY}\N),\qquad
\Gamma^0_{23}=-\frac{\p_\Sigma\N}{2\N},\qquad\quad
\\[2mm]
\Gamma^1_{33}&=&\frac{H_{\BY\BY}}{2},\qquad\Gamma^3_{33}=-\frac{\pb\N}{\N},
\eea
where
\be
\N=\left[1-\bigl(W\Sigma-\Y\BY+(\Sigma\theta(\Sigma)+
W\theta(W))G\bigr)/2a^2\right]^2.
\ee
At the point $W=0$, the tensor $h^{MN}$ can be represented as
\be
h^{MN}=\frac{1}{\det(g_{\alpha\beta})}\begin{pmatrix}
0\phm&0&\phm0&\phm0\\[1mm]
0\phm&-2\pp\Psi\pb\Psi\phm&\phm\pb\Psi&\phm\pp\Psi\,\\[1mm]
0\phm&\pb\Psi&\phm0&-1\\[1mm]
0\phm&\pp\Psi&-1&\phm0\end{pmatrix}.
\ee
Finally, the equation for the trapped surface can be found by direct
calculations:
\be
\left(\left(1+\frac{\Y\BY}{2a^2}\right)^2\p_{\Y\BY}+\frac{1}{a^2}\right)
\frac{2\Psi-H}{1+\Y\BY/2a^2}=0.
\ee

\end{document}